\documentclass[aps,preprint,preprintnumbers,amsmath,showpacs,amssymb,nofootinbib]{revtex4}
\usepackage{amssymb}
\usepackage{amsmath}
\usepackage{bm}
\draft
\begin{document}
\title{Analytical treatment of SUSY Quasi-normal modes in a non-rotating Schwarzschild black hole.}
\author{$^{1,2}$ Pablo Alejandro S\'anchez\footnote{pabsan@mdp.edu.ar}, $^{1,2}$ Jes\'us Mart\'{\i}n
Romero\footnote{jesusromero@conicet.gov.ar},
$^{1,2}$ Mauricio Bellini\footnote{mbellini@mdp.edu.ar}}
\address{$^{1}$ Departamento de F\'{\i}sica, Facultad de Ciencias Exactas y
Naturales, \\
Universidad Nacional de Mar del Plata, Funes 3350, (7600) Mar del
Plata, Argentina. \\ \\
$^{2}$ Instituto de Investigaciones F\'{\i}sicas de Mar del Plata
(IFIMAR), Consejo Nacional de Investigaciones Cient\'{\i}ficas y
T\'ecnicas (CONICET), Argentina.}
\begin{abstract}
We use the Fock-Ivanenko formalism to obtain the Dirac equation
which describes the interaction of a massless $1/2$-spin neutral
fermion with a gravitational field around a Schwarzschild black
hole (BH). We obtain approximated analytical solutions for the
eigenvalues of the energy (quasi-normal frequencies) and their
corresponding eigenstates (quasi-normal states). The interesting
result is that all the asymptotic states [and their supersymmetric
(SUSY) partners] have a purely imaginary frequency, which can be
expressed in terms of the Hawking temperature $T_H$:
$E^{(\uparrow\downarrow)}_n=-2\pi\,i\,n T_H$. Furthermore, as one
expects for SUSY Hamiltonians, the isolated bottom state has a
real null energy eigenvalue.
\end{abstract}
 \maketitle

\section{Introduction}

The BH can be understood as a thermodynamical system whose
(Hawking) temperature and entropy are given in terms of its global
characteristics (mass, charge and angular momentum). They are
obtained by solving a wave equation for small fluctuations subject
to the conditions that the flux be ingoing at the horizon and
outgoing at asymptotic infinity. Generally, these conditions lead
to a set of discrete complex eigenfrequencies, where the real part
represents the frequency of oscillation and the imaginary part
represents the damping. The frequencies and damping time of the
quasi-normal oscillations called "quasi-normal modes" (QNMs) are
determined only by the black hole's parameters and are independent
of the initial perturbations. It should be possible to infer the
black hole parameters solely from the QNM frequencies. QNMs carry
information of black holes and neutron stars, and thus are also of
great importance to gravitational-wave astronomy\cite{review}.
These oscillations are mainly produced during the formation phase
of compact stellar objects and can be strong enough to be detected
by several large gravitational wave detectors under construction.
Recently, QNMs of particles with different spins in black hole
spacetimes have also received much attention\cite{..}.

Numerical relativity has provided evidence that the QNMs dominate
the gravitational-wave signal associated with many processes
involving dynamical black holes (such as the formation of black
holes in gravitational collapse or binary merger). Since the QNMs
encode information concerning the parameters of the black hole one
may hope that the gravitational-wave detectors that are now coming
into operation will be able to use these signals to investigate
the black-hole population of the universe. Even though most
studies of QNMs have been motivated by their potential
astrophysical relevance, there are several other reasons why one
might be interested in understanding the spectrum of oscillations
of a black hole. In particular, the modes have played a role in
discussions of black-hole stability\cite{whi}. Recent
investigations of large extra dimensions, has led to the somewhat
striking prediction that mini black holes may be observed at
particle accelerators such as the LHC\cite{dimo}. One problem is
that in order to suppress a rapid proton decay we need to
physically split the quarks and leptons. Such models are
generically called split fermion models\cite{AH}. Most methods in
evaluating the QNMs are numerical. Recently, the Dirac field QNMs
were evaluated for a Schwarzschild black hole\cite{Cho}. A
powerful WKB scheme was devised by Schutz and Will\cite{SW}, and
it was also extended to higher orders\cite{IW}. In particular, the
interaction of fermions with a gravitational field is an important
topic which was firstly studied by Brill and Wheeler\cite{BW} in
the 1950s, and later by many other authors\cite{ot}. It is well
known that the vector field equation in a Schwarzschild spacetime
can be separated and the time and angular equation integrated. One
is then left with the separated radial equation whose integration
is not immediate. The remaining Schr\"odinger-like equation has
received great attention and provides QNMs.

On the other hand, the neutrino does not respond directly to
electric or magnetic fields. Therefore, if one wishes to influence
its orbit by forces subject to simple analysis, one has to make
use of gravitational fields. In other words, one has to consider
the physics of a neutrino in a curved metric. In this work we
revisit this topic using some ideas of the Fock-Ivanenko formalism
and SUSY to study the radiation of energy produced by a neutrino,
which is affected by a gravitational field produced by a
Schwarzschild BH. To do so it we shall study approximated
analytical solutions for the Dirac equation for a massless
$1/2$-spin-fermion around a neutral and non-rotating Schwarzschild
BH. This topic has been recently studied using numerical methods,
in a $d+1$ spacetime, where the $d-2$ extra dimensions are
compact\cite{uu}. In this paper we expand the formalism there
developed, making use of the supersymmetrical (SUSY) properties of
the Schr\"odinger equation.

The paper is organized as follows: in Sect. II we obtain the Dirac
equation for a $1/2$-spin massless fermion without charge
(neutrino) in a static and spherically symmetric spacetime, using
the Fock-Ivanenko formalism. In Sect. III we obtain the radial
equation for the spinors on such a metric. The spinors obey
Schr\"odinger-like stationary equations. In particular, in Sect.
IV we study these equations for a Schwarzschild BH. In Sect. V we
study the SUSY properties of the effective radial potentials of
the Schr\"odinger-like equations. In Sect. VI we obtain
approximated solutions of the quasi-normal frequencies and their
eigenstates and SUSY partners for the spinors. In both asymptotic
cases, he neutrinos are close, and very distant to the BH. The
quantization of quasi-normal modes is studied in the appendix. In
Sect. VII we have included some comments about the treatment of
$d>4$- neutrinos where the additional dimensions are compact.
Finally, in Sect. VIII we develop the final comments.

\section{Dirac equation for a neutrinos in a static and spherically symmetric spacetime}

We consider the metric that describes a static and spherically
symmetric spacetime
\begin{equation}\label{m1}
d\bar{s}^2 = f(r) dt^2 - f^{-1}(r)\, dr^2 + r^2\, d\Omega^2,
\end{equation}
where $\Omega$ is the solid angle, such that $d\Omega^2 =
\sin^2(\theta)\,d\theta^2 + \cos^2(\theta)\,d\phi^2 $. In this
paper we shall use natural units: $\hbar=c=1$. The Dirac equation
for a free massless fermion $\Psi$ with spin $1/2$, on the metric
(\ref{m1}), is
\begin{equation}
\bar\gamma^{\mu} \nabla_{\mu} \bar\Psi =0,
\end{equation}
where $\bar\nabla_{\mu}$ denotes the $\mu$th component of the
covariant derivative and $\bar\gamma^{\mu}$ are the Dirac matrix
components on the metric (\ref{m1}). Using conformal
transformations, one obtains
\begin{eqnarray}
\bar{g}_{\mu\nu} &\rightarrow & {g}_{\mu\nu} = \Theta^2
\,\bar{g}_{\mu\nu},
\\
\bar\Psi &\rightarrow & \Psi = \Theta^{-3/2} \bar\Psi, \\
\bar{\gamma}^{\mu} \bar{\nabla}_{\mu} \bar\Psi &\rightarrow &
\Theta^{5/2} \gamma^{\mu} \nabla_{\mu} \Psi,
\end{eqnarray}
where we shall take $\Theta=1/r$, so that $\Psi = r^{3/2}
\bar\Psi$. The resulting metric is
\begin{equation}\label{m2}
d{ s}^2 = \frac{f}{r^2} dt^2 - \frac{1}{f r^2} dr^2 - d\Omega^2,
\end{equation}
on which
\begin{equation}\label{d2}
\gamma_{\mu} \nabla^{\mu} \Psi=0.
\end{equation}
The spinors $\chi^{(\uparrow\downarrow)}(\theta,\phi)$ on the
$2$-sphere obey the equation\footnote{Notation: in what follows we
shall use the compact notation denoting by $\uparrow $ all that
correspond with the spinor "up", and with $\downarrow $ all that
correspond with the spinor "down". For instance, the scalar
$\kappa_{\uparrow}=l+1$ corresponds to $\chi^{(\uparrow)}$ and
$\kappa_{\downarrow}=-l$ corresponds to
$\chi^{(\downarrow)}$.}\cite{landau}
\begin{equation}
\gamma_i \nabla^i \chi_l^{(\uparrow\downarrow)} =  i \,\left( \begin{array}{ll} (l+1) \\
(-l)
\end{array} \right) \chi_l^{(\uparrow\downarrow)}=  i \,(\kappa_{\uparrow\downarrow}) \,\chi_l^{(\uparrow\downarrow)},
\end{equation}
where
\begin{equation}\label{ka}
\kappa_{\uparrow\downarrow} =
\left( \begin{array}{ll} l+1 \\
-l \end{array} \right).
\end{equation}
Notice that $(\kappa_{\uparrow\downarrow})$ can take integer
values, but the zero is excluded: $(\kappa_{\uparrow\downarrow})
\neq 0$. If we expand the function $\Psi$ as
\begin{equation}
\Psi = \sum_{l} \left( \phi_l^{(\uparrow)} \chi_l^{(\uparrow)} +
\phi_l^{(\downarrow)}\chi_l^{(\downarrow)}\right),
\end{equation}
the equations of motion for $\phi_l^{(\uparrow)}$ and
$\phi^{(\downarrow)}$ as a result are found to be\cite{CH}
\begin{eqnarray}
\left\{ \gamma^0 \nabla_0 + \gamma^1 \nabla_1 + i\,\left(
l+1\right) \gamma^5  \right\} \phi_l^{(\uparrow)} &=& 0, \quad l=0,1,..., \label{equa1}\\
\left\{ \gamma^0 \nabla_0 + \gamma^1 \nabla_1 - i\, l\,\gamma^5
l\right\} \phi_l^{(\downarrow)}& = & 0, \quad l=1,2,...,
\label{equa2}
\end{eqnarray}
where we use the spatiotemporal coordinates $(t,r,\theta,\phi)$
and $\gamma^5=\gamma^0 \gamma^1 \gamma^2 \gamma^3 $ is the
interaction. With the aim to solve the equations (\ref{equa1}) and
(\ref{equa2}), we shall make use of the Fock-Ivanenko
formalism\cite{fi}. To represent the system we choose the
following Dirac matrices:
\begin{equation}\label{gamma}
\gamma^t = -i\,\frac{r}{\sqrt{f(r)}}\ \sigma^3, \qquad \gamma^r =
r\,\sqrt{f(r)}\, \sigma^2.
\end{equation}
where the $\sigma^i$ are the Pauli matrices
\begin{equation}
\sigma^1 = \left(\begin{array}{ll} 0 \qquad 1 \\ 1 \qquad 0
\end{array}\right), \qquad \sigma^2 = \left(\begin{array}{ll} 0 \qquad -i \\ i \qquad 0
\end{array}\right), \qquad \sigma^3 = \left(\begin{array}{ll} 1 \qquad 0 \\ 0 \qquad
-1 \end{array}\right),
\end{equation}
so that the spin connections, $\Gamma_{\mu} = {1\over 4}
\gamma^{\nu\mu} {\partial \over \partial x^{\nu}}
g_{\mu\mu}$\footnote{The reader can see, for example, page 113 -
chapter 6 - of the book \cite{fi}.}, are given by\footnote{Notice
that the sum does not run over $\mu$, but it does over $\nu$.}
\begin{equation}
\Gamma_t = \sigma^1 \left(\frac{r^2}{4}\right) \frac{d}{dr}
\left(\frac{f(r)}{r^2}\right), \qquad \Gamma_r =0.
\end{equation}

\section{Equations of motion for $\phi_l^{(\uparrow\downarrow)}$}

\subsection{Spinor $\phi_l^{(\uparrow)}$}

The equation of motion for $\phi_l^{(\uparrow)}$ (the equation for
$\phi_l^{(\downarrow)}$ can be worked out in the same way) is
\begin{equation}\label{spinor}
\sigma_2\, r \sqrt{f(r)} \left[ \frac{\partial }{\partial r} +
\frac{r}{2\sqrt{f(r)}} \frac{d}{dr} \left(
\frac{\sqrt{f(r)}}{r}\right) \right] \phi_l^{(\uparrow)} - i
\sigma_1 \left( l +1\right) \phi_l^{(\uparrow)} = i \sigma_3
\left(\frac{r}{\sqrt{f(r)}}\right) \frac{\partial
\phi_l^{(\uparrow)}}{\partial t}.
\end{equation}
We propose solutions of the form \begin{equation}
\phi_l^{(\uparrow)}(r,t) =
\left(\frac{\sqrt{f(r)}}{r}\right)^{-1/2}
e^{-i E^{(\uparrow)}_n t} \left( \begin{array}{ll} i\, \Psi_1^{(\uparrow)}(r) \\
\Psi_2^{(\uparrow)}(r)
\end{array} \right),
\end{equation}
where $E^{(\uparrow)}_n$ is some constant of integration. With
this transformation, the Dirac equation can be written as two
coupled first-order equations:
\begin{eqnarray}
f(r)\,\frac{d \Psi_1^{(\uparrow)}}{dr} - \frac{\sqrt{f(r)}}{r} \left( l+1 \right) \Psi_1^{(\uparrow)}(r)& = & E^{(\uparrow)}_n\, \Psi_2^{(\uparrow)}(r) ,  \label{1a}\\
f(r)\,\frac{d \Psi_2^{(\uparrow)}}{dr}+\frac{\sqrt{f(r)}}{r}
\left( l+1 \right) \Psi_2^{(\uparrow)}(r)& = & -E^{(\uparrow)}_n\,
\Psi_1^{(\uparrow)}(r). \label{1b}
\end{eqnarray}
Now we define the tortoise coordinate $u$ and the function $W(r)$,
as
\begin{equation}\label{tor}
f(r) \frac{d}{dr} = \frac{d}{du}, \qquad W^{(\uparrow)}(r) =
\frac{(l+1)}{r} \sqrt{f(r)},
\end{equation}
so that the equations (\ref{1a}) and (\ref{1b}) can be written,
respectively, as
\begin{eqnarray}
&& \left[\frac{d}{du} - W^{(\uparrow)}(u) \right] \Psi_1^{(\uparrow)}(u) = E^{(\uparrow)}\, \Psi_2^{(\uparrow)}(u),  \label{eu1}\\
&& \left[\frac{d}{du} + W^{(\uparrow)}(u) \right]
\Psi_2^{(\uparrow)}(u) = -E^{(\uparrow)}\, \Psi_1^{(\uparrow)}(u).
\label{eu2}
\end{eqnarray}
These equations can be separated as
\begin{eqnarray}
&& \left(-\frac{d^2}{du^2} + V^{(\uparrow)}_1(u)\right) \Psi_1^{(\uparrow)}(u) = \left[E^{(\uparrow)}_n\right]^2 \Psi_1^{(\uparrow)}(u), \label{s1} \\
&& \left( - \frac{d^2}{du^2} + V^{(\uparrow)}_2(u)\right)
\Psi_2^{(\uparrow)}(u) = \left[E^{(\uparrow)}_n\right]^2
\Psi_2^{(\uparrow)}(u), \label{s2}
\end{eqnarray}
such that the supersymmetric (SUSY) potentials $V^{(\uparrow)}_1$
and $V^{(\uparrow)}_2$ are given by the expression
\begin{equation}\label{pot}
V^{(\uparrow)}_{1,2}(u) = \mp \frac{dW^{(\uparrow)}}{du} +
\left[W^{(\uparrow)}(u)\right]^2.
\end{equation}
The effective potentials in (\ref{pot}) are supersymmetric. Hence,
the functions $\Psi_1^{(\uparrow)}$ and $\Psi_2^{(\uparrow)}$ have
the same spectrum for quasi-normal modes.

\subsection{Spinor $\phi_l^{(\downarrow)}$}

In the same manner one can obtain the Schr\"odinger-like equations
once we propose the solutions
\begin{equation}
\phi_l^{(\downarrow)}(r,t) =
\left(\frac{\sqrt{f(r)}}{r}\right)^{-1/2}
e^{-i E^{(\downarrow)}_n t} \left( \begin{array}{ll} i\, \Psi_1^{(\downarrow)}(r) \\
\Psi_2^{(\downarrow)}(r)
\end{array} \right),
\end{equation}
such that the superpotential $W^{(\downarrow)}(r)$ is
\begin{equation}
W^{(\downarrow)}(r) = -\frac{l}{r} \sqrt{f(r)},
\end{equation}
and the Schr\"odinger-like equations for $\Psi_1^{(\downarrow)}$
$\Psi_2^{(\downarrow)}$ are
\begin{eqnarray}
&& \left(-\frac{d^2}{du^2} + V^{(\downarrow)}_1(u)\right) \Psi_1^{(\downarrow)}(u) = \left[E^{(\downarrow)}_n\right]^2 \Psi_1^{(\downarrow)}(u), \label{ss1} \\
&& \left( - \frac{d^2}{du^2} + V^{(\downarrow)}_2(u)\right)
\Psi_2^{(\downarrow)}(u) = \left[E^{(\downarrow)}_n\right]^2
\Psi_2^{(\downarrow)}(u), \label{ss2}
\end{eqnarray}
with the potentials $V^{(\downarrow)}_{1,2}(u)$ given by
\begin{equation}\label{pot1}
V^{(\downarrow)}_{1,2}(u) = \mp \frac{dW^{(\downarrow)}}{du} +
\left[W^{(\downarrow)}(u)\right]^2.
\end{equation}
The equations (\ref{s1}-\ref{s2}) and (\ref{ss1}-\ref{ss2}) give
us the radial information about the wave functions of the spinors
$\phi^{(\uparrow,\downarrow )}_l(r,t)$. In the following sections
we shall consider an analytical treatment to solve the pairs of
Schr\"odinger-like equations (\ref{s1}-\ref{s2}) and
(\ref{ss1}-\ref{ss2}), for a Schwarzschild black hole.

\section{Neutrinos in a Schwarzschild black-hole}

We consider the particular case where the function $f(r)$ in the
metric (\ref{m1}) is given by
\begin{equation}
f(r)= 1 -\frac{a}{r},
\end{equation}
such that $a=2 G M$ is the Schwarzschild radius. In this case the
tortoise coordinate $u(r)$ is given by
\begin{equation}\label{u}
u(r)= r + a\, {\rm ln}\left[\frac{r}{a} -1\right], \qquad {\rm
or}, \qquad  u(r)= \frac{a}{1-f(r)} + a\,{\rm
ln}\left[\frac{f(r)}{1-f(r)}\right].
\end{equation}
From the second equation in (\ref{u}), we obtain
\begin{equation}\label{f}
\left[1-f(u)\right] = \frac{1}{1 + LW\left(e^{(u-a)/a}\right)},
\end{equation}
where $LW(x)$ is the Lambert function, such that $LW(x)\,
e^{LW(x)}=x$ and $LW(0)=0$. Now we are aimed to obtain the
potentials $V^{(\pm)}_{1,2}(u)$ as a function of $u$. From eqs.
(\ref{tor}), the second equation of (\ref{u}) and (\ref{f}), we
obtain the superpotentials
\begin{equation}
W^{(\uparrow\downarrow)}(u) =
\frac{(\kappa_{\uparrow\downarrow})}{a}
\frac{\left[LW\left[e^{(u-a)/a}\right]\right]^{1/2}}{\left[1+LW\left[e^{(u-a)/a}\right]\right]^{3/2}}.
\end{equation}
The exact potentials, written as a function of $u$, are
\begin{eqnarray}
V^{(\uparrow\downarrow)}_{1,2}(u) & = &
\left[W^{(\uparrow\downarrow)}\right]^2(u) \pm
\frac{d\,W^{(\uparrow\downarrow)}(u)}{du} =
\frac{\left(\kappa_{\uparrow\downarrow}\right)^2}{a^2}
\frac{LW\left[e^{(u-a)/a}\right]}{\left[1+LW\left[e^{(u-a)/a}\right]\right]^{3}}
\nonumber \\
& \mp & \frac{1}{2} \frac{(\kappa_{\uparrow\downarrow})}{a^2}
\frac{\left[2\,LW\left[e^{(u-a)/a}\right]-1\right]}{\left[1+LW\left[e^{(u-a)/a}\right]\right]^{7/2}}
\sqrt{LW\left[e^{(u-a)/a}\right]}.\label{potet}
\end{eqnarray}
In the limit case where the particle is close to the Schwarzschild
radius: $u \rightarrow -\infty$, so that $e^{(u-a)/a} \rightarrow
0$. In this case the effective potentials can be approximated by
\begin{equation}\label{po}
V^{(\uparrow\downarrow)}_{1,2}(u)  \simeq
\frac{\left(\kappa_{\uparrow\downarrow}\right)^2}{a^2} e^{(u-a)/a}
\pm \frac{(\kappa_{\uparrow\downarrow})}{2 a^2} e^{(u-a)/(2a)}.
\end{equation}
In order to make a dimensionless description of the problem we can
make the following transformation: $ 2v= - (u-a)/a$, and the
Schr\"odinger-like equations (\ref{s1}) and (\ref{s2}), for
$\Psi_1^{(\uparrow\downarrow)}(v)$ and
$\Psi_2^{(\uparrow\downarrow)}(v)$, close the Schwarzschild
radius, become
\begin{equation}\label{sch}
\left\{-\frac{d^2}{dv^2} + 4
\left(\kappa_{\uparrow\downarrow}\right)^2 e^{-2 v} \pm 2
\left(\kappa_{\uparrow\downarrow}\right) \,e^{-v}
 - 4 a^2\, \left(\begin{array}{ll}  \left[E^{(i,\uparrow\downarrow)}_{n}\right]^2
\\  \left[E^{(j,\uparrow\downarrow)}_n\right]^2  \end{array} \right)
\right\} \left(\begin{array}{ll}
\Psi_i^{(\uparrow\downarrow)}(\alpha,l,v)
\\ \Psi_j^{(\uparrow\downarrow)}(\beta ,l,v)  \end{array} \right)=0,
\end{equation}
where $E^{(i,\uparrow\downarrow)}_{n}=
E^{(j,\uparrow\downarrow)}_{n}$ and
\begin{eqnarray}
i& = & 1, \quad j=2 \quad \alpha=n+1, \quad \beta=n,\qquad {\rm for} \qquad \phi^{(\uparrow)}_l, \\
i& = & 2, \quad j=1 \quad \alpha=n, \quad \beta=n+1, \qquad {\rm
for} \qquad \phi^{(\downarrow)}_l.
\end{eqnarray}
Notice that for large $r$ one obtains $u \rightarrow -\infty$, so
that $V^{(\uparrow\downarrow)}_{1,2} \rightarrow 0$. In other
words, the approximated potential (\ref{po}) becomes null for
large distances and takes the same asymptotic value as the exact
potential (\ref{potet}). Therefore, the asymptotic solutions of
the eq. (\ref{sch}) will be a good approximation for $r\rightarrow
\infty$. However, the approximated potential (\ref{po}) is not a
good approximation for intermediate distances.

\section{Supersymmetry of Hamiltonians}

We consider the Hamiltonians ${\rm H}^{(\uparrow\downarrow)}_1$
and ${\rm H}^{(\uparrow\downarrow)}_2$
\begin{eqnarray}
{\rm H}^{(\uparrow\downarrow)}_1 &=& -\frac{d^2}{dv^2} + \left[W^{(\uparrow\downarrow)}\right]^2(v)
- \frac{dW^{(\uparrow\downarrow)}(v)}{dv}, \label{h1}\\
{\rm H}^{(\uparrow\downarrow)}_2 &=& -\frac{d^2}{dv^2} +
\left[W^{(\uparrow\downarrow)}\right]^2(v) +
\frac{dW^{(\uparrow\downarrow)}(v)}{dv}, \label{h2}
\end{eqnarray}
such that ${\rm H}^{(\uparrow\downarrow)}_2-{\rm
H}^{(\uparrow\downarrow)}_1 = 2 {dW^{(\uparrow\downarrow)}(v)
\over dv}$. In our case $W^{(\uparrow\downarrow)}(v) \simeq -2
(\kappa_{\uparrow\downarrow}) e^{-v}$\footnote{Actually, there
exist a family of solutions $W^{(\uparrow\downarrow)}(v) = {C [
2(\kappa_{\uparrow\downarrow}) e^{-v}\,{\rm
Ei}(1,-4(\kappa_{\uparrow\downarrow}) e^{-v}) +
2(\kappa_{\uparrow\downarrow}) e^{-v} \over C\,{\rm
Ei}(1,-4(\kappa_{\uparrow\downarrow}) e^{-v}) +1}$, for ${\rm
Ei}(az) = z^{a-1}\,\Gamma(1-a,z)$, which give us SUSY potentials
$V^{(\uparrow\downarrow)}_{1,2}(v) =
4(\kappa_{\uparrow\downarrow})^2 e^{-2v} \pm
2(\kappa_{\uparrow\downarrow}) e^{-v}$, but in this paper we shall
restrict our study to the particular solution with $C=0$.}, so
that we obtain ${\rm H}^{(\uparrow\downarrow)}_1-{\rm
H}^{(\uparrow\downarrow)}_2 =  2 W^{(\uparrow\downarrow)}(v)$.
Following Mielnik\cite{miel}, we shall define the differential
operators:
\begin{eqnarray}
{\rm b^*_{(\uparrow\downarrow)}} & = &  -\frac{d}{dv} + W^{(\uparrow\downarrow)}(v), \\
{\rm b_{(\uparrow\downarrow)}} & = & \frac{d}{dv} +
W^{(\uparrow\downarrow)}(v),
\end{eqnarray}
then, we see that Hamiltonians (\ref{h1}) and (\ref{h2}) can be
rewritten in terms of ${\rm b_{(\uparrow\downarrow)}}$ and ${\rm
b^*_{(\uparrow\downarrow)}}$:
\begin{equation}
{\rm b^*_{(\uparrow\downarrow)}} {\rm b_{(\uparrow\downarrow)}}=
{\rm b_{(\uparrow\downarrow)}} {\rm b^*_{(\uparrow\downarrow)}} +
[{\rm b^*_{(\uparrow\downarrow)}}, {\rm b_{(\uparrow\downarrow)}}]
= {\rm b_{(\uparrow\downarrow)}} {\rm b^*_{(\uparrow\downarrow)}}
+ 2 W^{(\uparrow\downarrow)}(v) = {\rm H}^{(\uparrow\downarrow)}_2
+ 2 W^{(\uparrow\downarrow)}(v) = {\rm
H}^{(^{(\uparrow\downarrow)})}_1.
\end{equation}
Furthermore, ${\rm H}^{(\uparrow\downarrow)}_i {\rm
b^*_{(\uparrow\downarrow)}} = {\rm b^*_{(\uparrow\downarrow)}}
{\rm b_{(\uparrow\downarrow)}} {\rm b^*_{(\uparrow\downarrow)}} =
{\rm b^*_{(\uparrow\downarrow)}} {\rm
H}^{(\uparrow\downarrow)}_j$, which means that
\begin{equation}
{\rm H}^{(\uparrow\downarrow)}_i \left( {\rm
b^*_{(\uparrow\downarrow)}} \Psi_{j}^{(\uparrow\downarrow)}\right)
= {\rm b^*_{(\uparrow\downarrow)}} \left( {\rm
H}^{(\uparrow\downarrow)}_j \Psi_{j}^{(\uparrow\downarrow)}
\right) = {\rm b^*_{(\uparrow\downarrow)}} \left(
\left[E^{(\uparrow\downarrow)}_n\right]^2
\Psi_{j}^{(\uparrow\downarrow)} \right) =
\left[E^{(\uparrow\downarrow)}_n\right]^2 \left({\rm
b^*_{(\uparrow\downarrow)}}
\Psi_{j}^{(\uparrow\downarrow)}\right).
\end{equation}
Therefore, if $\Psi_{j}^{(\uparrow\downarrow)}(\alpha,l,v)$ is and
eigenvector of ${\rm H}^{(\uparrow\downarrow)}_j$ with eigenvalue
$\left[E^{(\uparrow\downarrow)}_n\right]^2$, hence
$\Psi_{i}^{(\uparrow\downarrow)}(\alpha,l,v)={\rm
b^*_{(\uparrow\downarrow)}}
\Psi_{j}^{(\uparrow\downarrow)}(\beta,l,v)$ will be an eigenvector
of ${\rm H}^{(\uparrow\downarrow)}_i$ with eigenvalue
$\left[E^{(\uparrow\downarrow)}_n\right]^2$. In other words the
potentials $V^{(\uparrow\downarrow)}_i$ and
$V^{(\uparrow\downarrow)}_j$ possess the same spectra of
quasi-normal frequencies (but with different eigenstates), because
they are supersymmetric partners derived from the same
superpotentials $W^{(\uparrow\downarrow)}(v)$. Notice that in
cases where $E^{(\uparrow\downarrow)}_n$ is not real, the
Hamiltonians will not be hermitian.

\section{Approximated solutions}

In order to solve the equations (\ref{sch}) we must consider that
such equations are only valid close to the Schwarzschild horizon:
$r\geq a$. In this limit, $v\rightarrow \infty$. We shall consider
separately the cases for the potentials
$V^{(\uparrow\downarrow)}_1$ and $V^{(\uparrow\downarrow)}_2$, for
the spinors $\phi^{(\uparrow\downarrow)}_l$.

\subsection{Spinor $\phi_l^{(\uparrow)}$}

As a first case we consider the radial equations for the spinor
$\phi_l^{(\uparrow)}$. We shall consider, separately, the
attractive and repulsive potentials in the equation (\ref{pot}).

\subsubsection{Attractive potential for $\phi_l^{(\uparrow)}$}

The first case consists in
\begin{equation}
V^{(\uparrow)}_2(l,v) =4 (l+1)^2 e^{-2 v} - 2 (l+1) \,e^{-v},
\end{equation}
which is similar (but not exactly equal) to a Morse
potential\footnote{It is interesting notice that the original
bound state Morse potential has the form
$V_{Morse}(v)=S^2\left[e^{-2v}-e^{-v}\right]$ and corresponds to
an Hermitian exactly solvable potential with real eigenvalues of
energy. However, as was demonstrated in \cite{mor}, it is possible
to extend the usual theory for exactly solvable (ES)\cite{es} and
quasi-exactly solvable\cite{qes} (QES) potentials to accommodate
QNMs solutions.}. The general solution for (\ref{sch}) is given by
\begin{equation}\label{solu}
\Psi_{2}^{(\uparrow)}(n,l,v) \simeq e^{-v/2} \left[  {\cal
I}_{\nu_1}[z(v)] -{\cal I}_{\nu_2}[z(v)]\right] \left[ B_{nl} +
C_{nl} \int\frac{dv \,\, e^{v}}{\left[  {\cal I}_{\nu_1}[z(v)]-
{\cal I}_{\nu_2}[z(v)]\right]^2}\right],
\end{equation}
where the ${\cal I}_{\nu_1}[z(v)]$ and ${\cal I}_{\nu_2}[z(v)]$
are the modified Bessel functions of the first kind, with
\begin{eqnarray}
&& \nu_{1,2} = 2\,i\,a \,E^{(\uparrow)}_n \mp \frac{1}{2}, \\
&& z(v) = 2 (l+1) \,\,e^{-v}.
\end{eqnarray}
As can be demonstrated [see appendix (\ref{ap1})] the eigenvalues
of energy are\footnote{In general the energy can take the values
$E^{(\uparrow)}_n=\mp i n/(2a)$, but we choose only the
eigenvalues with sign minus because they are those that correspond
to decaying modes.}:
\begin{equation}
E^{(2,\uparrow)}_n = -i\, \frac{n}{2 a}, \qquad n=1,2, ...
\end{equation}
and the eigenvalues for the angular moment are $l\geq 0$. In order
to avoid a possible divergence in (\ref{solu}), we choose
$C_{nl}=0$. A particular manifestation of Hamiltonians which are
SUSY, is related to the null eigenvalue of energy:
$E^{(\uparrow)}_{n=0}=0$. In this case $n=0$ and the solution is
\begin{equation}\label{es1}
\Psi_{2}^{(\uparrow)}(0,l,v) \simeq B_{0l}\, e^{-v/2} \left[
{\cal I}_{-1/2}[z(v)] - \,{\cal I}_{1/2}[z(v)]\right] \propto  \,
e^{- 2(l+1)\, e^{-v}},
\end{equation}
where $B_{0l}$ is a constant to be determined by normalization.

The $\Psi^{(\uparrow)}_{2}(n,l,v)$-partner states can be obtained
when we apply the operator of $b^*_{\uparrow}$ to
$\Psi^{(\uparrow)}_{2}(n,l,v)$
\begin{equation}\label{susy}
\Psi_{1}^{(\uparrow)}(n+1,l,v)=\left\{- \frac{d}{dv} +
2(l+1)\,e^{-v} \right\} \,\Psi_{2}^{(\uparrow)}(n,l,v).
\end{equation}
The correspondence (\ref{susy}) is a manifestation of the SUSY
character of the Hamiltonians ${\rm H}^{(\uparrow)}_1$ and ${\rm
H}^{(\uparrow)}_2$.

\subsubsection{Repulsive potential for $\phi_l^{(\uparrow)}$}

Now we consider the potential
\begin{equation}
V^{(\uparrow)}_1(v) =4 (l+1)^2 e^{-2 v} + 2 (l+1) \,e^{-v}.
\end{equation}
In this case the general solution is given by
\begin{eqnarray}
\Psi_{1}^{(\uparrow)}(n & + & 1,l,v)\simeq  \bar{B}_{nl} \left\{
\left[ \left(l+1\right) e^{-v/2} + \left(n+{1\over 2}\right)
e^{v/2} \right]  {\cal I}_{\mu_1}[z(v)] + e^{-v/2} \left(
l+1\right) {\cal
I}_{\mu_2}[z(v)]\right\} \nonumber \\
& + & \bar{C}_{nl} \left\{ \left[\left( (n+1) 4^{n+1} 16^{-(n+1)}
- \frac{4^{-(n+1)}}{2} \right) e^{v/2} + e^{-v/2} \left(
l+1\right) 4^{(n+1)} 16^{-(n+1)} \right] {\cal I}_{\mu_1}[z(v)]
\right.
\nonumber \\
&+ & \left. e^{-v/2}
4^{(n+1)} 16^{-(n+1)} \left(l+1\right) {\cal I}_{\mu_2}[z(v)]\right\} \nonumber \\
& \times & \frac{1}{4} \int \frac{dv\, e^{v}}{\left\{ \left[ l+1 +
\left(n+\frac{1}{2}\right) e^{v} \right] {\cal I}_{\mu_1}[z(v)] +
(l+1) {\cal I}_{\mu_2}[z(v)]\right\}^2}, \label{solu1}
\end{eqnarray}
where
\begin{eqnarray}
&& \mu_{1} = n+1/2, \qquad \mu_2= n+3/2, \\
&& z(v) = 2 (l+1) \,\,e^{-v}.
\end{eqnarray}
In order to the SUSY expression for the partners (\ref{susy}) to
be fulfilled, we shall require that $\bar{B}_{nl}=0$, so that the
resulting nonzero constants $B_{nl}$ in (\ref{solu}) and
$\bar{C}_{nl}$ in (\ref{solu1}) should be determined by
normalization. As can be demonstrated [see appendix (\ref{ap1})],
the energy eigenvalues are
\begin{equation}
E^{(1,\uparrow)}_{n} = -i\, \frac{(n+1)-1}{2 a}= -i\, \frac{n}{2
a}, \qquad n=1,2, ...\,.
\end{equation}
Notice that $E^{(2,\uparrow)}_{n}=E^{(1,\uparrow)}_{n}$ [see eq.
(\ref{sch})].

\subsection{Spinor $\phi_l^{(\downarrow)}$}

For completeness, we consider the radial equations for the spinor
$\phi_l^{(\downarrow)}$, such that the attractive and repulsive
potentials are given by (\ref{pot1}).

\subsubsection{Attractive potential for $\phi_l^{(\downarrow)}$}

Now we consider the attractive potential related to
$\kappa^{(\downarrow)}=-l$. In this case the potential is
\begin{equation}
V^{(\downarrow)}_1(v) =4 \,l^2 e^{-2 v} - 2\, l \,e^{-v},
\end{equation}
so that the general solution of (\ref{sch}) is
\begin{eqnarray}
\Psi_{1}^{(\downarrow)}(n,l,v) &\simeq & e^{-v/2} \, \left\{
\alpha_{nl}\,\left[ {\cal I}_{n-1/2}\left[-2\l\,e^{-v}\right] +
{\cal I}_{n+1/2}\left[-2\l\,e^{-v}\right]\right] \right. \\
&+& \left. \beta_{nl}\,\left[ {\cal
K}_{n-1/2}\left[-2\l\,e^{-v}\right] - {\cal
K}_{n+1/2}\left[-2\l\,e^{-v}\right]\right]\right\},
\end{eqnarray}
such that the functions ${\cal K}_{n\mp 1/2}$ are the modified
Bessel functions of second kind, and the quasi-normal frequencies
are
\begin{equation}
E^{(1,\downarrow)}_n = - i \frac{n}{2a}, \qquad n=0,1,...\,.
\end{equation}
The particular case with $n=0$ give us the bottom energy
eigenvalue $E^{(1,\downarrow)}_{n=0}=0$, which corresponds to the
eigenfunction
\begin{equation}
\Psi_{1}^{(\downarrow)}(0,l,v) = \alpha_{0l}\,e^{-2l\,e^{-v}},
\end{equation}
where we have put $\beta_{nl}=0$, for the eigenfunction to be
finite along all the domain $v$.

\subsubsection{Repulsive potential for $\phi_l^{(\downarrow)}$}

Finally, for completeness, we consider the repulsive potential
associated to $\kappa^{(\downarrow)}=-l$. The potential has the
form
\begin{equation}
V^{(\downarrow)}_2(v) =4 \,l^2 e^{-2 v} + 2\, l \,e^{-v},
\end{equation}
and the general solution in this case is
\begin{eqnarray}
\Psi_{2}^{(\downarrow)}(n+1,l,v) &\simeq & e^{-v/2} \, \left\{
\bar{\alpha}_{nl}\,\left[ {\cal I}_{n-1/2}\left[2\l\,e^{-v}\right]
+
{\cal I}_{n+1/2}\left[2\l\,e^{-v}\right]\right] \right. \nonumber \\
&+& \left. \bar{\beta}_{nl}\,\left[ {\cal
K}_{n-1/2}\left[2\l\,e^{-v}\right] - {\cal
K}_{n+1/2}\left[2\l\,e^{-v}\right]\right]\right\}. \label{soluu}
\end{eqnarray}
Finally, the quasi-normal frequencies are
\begin{equation}
E^{(2,\downarrow)}_n = - i \frac{n}{2a} \qquad n=1,2,...\,.
\end{equation}
In order to fulfill the SUSY expression for the partners
\begin{equation}\label{susy1}
\Psi_{2}^{(\downarrow)}(1,l,v)=\left\{- \frac{d}{dv} -
2\,l\,e^{-v} \right\} \,\Psi_{1}^{(\downarrow)}(0,l,v),
\end{equation}
we shall require that $\bar{\alpha}_{0l}=0$ and more generally,
that $\bar{\alpha}_{nl}=0$ in (\ref{soluu}).

\subsection{Asymptotic solutions for large distances: $r\rightarrow \infty$}

To complete our study we shall study the weak gravitational field
case for large distances to the BH. In this case
$r\rightarrow\infty$, so that $e^{(u-a)/a} \rightarrow e^{u/a}$.
The superpotential can be approximated by
\begin{equation}
\left.W^{(\uparrow\downarrow)}(u)\right|_{u\rightarrow\infty}
\simeq \frac{(\kappa_{\uparrow\downarrow})}{a}
\frac{1}{LW[e^{u/a}]},
\end{equation}
so that the SUSY potentials will be
\begin{equation}
\left.V^{(\uparrow\downarrow)}_{1,2}(u)\right|_{u\rightarrow
\infty} \simeq \frac{(\kappa_{\uparrow\downarrow})}{a^2}
\frac{1}{\left\{LW[e^{u/a}]\right\}^2}
\left[(\kappa_{\uparrow\downarrow}) \mp 1\right].
\end{equation}
For very large distances these potentials tend to $0$:
$\left.V^{(\uparrow\downarrow)}_{1,2}(u)\right|_{u\rightarrow
\infty} \rightarrow 0$, in agreement with the approximated
potential (\ref{po}). Hence, the approximated Schr\"odinger-like
equations for $u\rightarrow \infty$ can be written as
\begin{equation}
-\frac{d^2}{du^2} \Psi_{1,2}^{(\uparrow\downarrow)}(u) \simeq
\left[E^{(\uparrow\downarrow)}_n\right]^2
\Psi_{1,2}^{(\uparrow\downarrow)}(u),
\end{equation}
which has a general solution that can be written as a linear
combination of $e^{\pm i E^{(\uparrow\downarrow)}_n u}$. After
taking into account the normalization conditions, and the
signature of $E^{(\uparrow\downarrow)}_n$, we obtain that for
large distances the outgoing solutions are
\begin{equation}
\Psi_{1,2}^{(\uparrow\downarrow)}(u) \sim e^{-nu/(2a)}.
\end{equation}
This solution agrees perfectly with the whole obtained in
\cite{cho}.

\subsection{Quasi-normal frequencies and Hawking temperature}

The Hawking temperature for the Schwarzschild BH is
\begin{equation}
T_H=\frac{1}{4\pi a},
\end{equation}
so that the quasi-normal frequencies can be written as
\begin{equation}
E^{(2,\uparrow\downarrow)}_{n}=E^{(1,\uparrow\downarrow)}_{n} = -
2\pi\,i\,n\,T_H.
\end{equation}
This result is exactly whole obtained recently in\cite{uu}, but
using the WKB method (see also\cite{cqg2004}).

\section{Neutrinos in a $d >4$-dimensional Schwarzschild BH}

The study of neutrinos which are close to multidimensional
Schwarzschild BH is an interesting issue. In this case the
extended $d >4$-dimensional Schwarzschild metric is given
by\cite{kon}
\begin{equation}\label{d}
d\bar{s}^2 = f_{(d)}(r) dt^2 - f^{-1}_{(d)} (r)\, dr^2 + r^2\,
d\Omega^2_{d-2},
\end{equation}
where $d$ is the dimension of the spacetime, $d\Omega^2_{d-2}$
denotes a metric of a $(d-2)$-dimensional sphere and
\begin{equation}
f_{(d)}(r) = 1 - \left(\frac{a}{r}\right)^{(d-3)}.
\end{equation}
In this case the scalars analogous to (\ref{ka}) are
\begin{equation}\label{ka1}
\kappa^{(d)}_{\uparrow\downarrow} =
\left( \begin{array}{ll} l+\left(\frac{d-2}{2}\right) \\
-\left[l+ \left(\frac{d-4}{2}\right)\right]\end{array} \right),
\end{equation}
where $l\geq -(d-2)/2+1$ and  $l\geq -(d-2)/2+2$ for
$\kappa^{(d)}_{\uparrow}$ and $\kappa^{(d)}_{\downarrow}$,
respectively. Note that $d$ only can take integer-even values in
order for $\kappa^{(d)}_{\uparrow\downarrow}$ to be a nonzero
integer. The tortoise coordinate can be written as
\begin{equation}
u(r) = a \,\left[\frac{1}{1-f_{(d)}(r)}\right]^{1/(d-3)} \,\,_2
{\cal F}_1\left\{1,\frac{1}{3-d};\frac{d-4}{d-3};\left[
\left[\frac{1}{1-f_{(d)}(r)}\right]^{1/(d-3)}-1\right]^{(3-d)}\right\},
\end{equation}
where $_2 {\cal F}_1[\mu,\nu;\gamma;x]=\sum_{n=0}^{\infty} {\nu_n
\mu_n x^n\over \gamma_n n!}$ is the hypergeometric function and we
have used the fact that
\begin{equation}
\left(a/r\right)= \left(\frac{1}{1-f_{(d)}(r)}\right)^{1/(d-3)}.
\end{equation}
In this case one can approximate the superpotential for very large
extra dimensions, by
\begin{equation}
\left.W^{(\uparrow\downarrow(d))}(r)\right|_{d\rightarrow\infty}
\simeq  \frac{(\kappa^{(d)}_{\uparrow\downarrow})}{a
\left[1+\left(\frac{u}{a}-1\right)^{1/(3-d)}\right]}
\left[1-\left[ 1+\left(\frac{u}{a}-1\right)^{1/(3-d)}
\right]^{(3-d)}\right]^{1/2},
\end{equation}
The detailed analysis in the SUSY framework is very complicated
and goes beyond the scope of this work, but finally one can find
that the eigenvalues of the energies are\cite{cho}
\begin{equation}
E^{(\uparrow\downarrow(d))}_n = -i \left(\frac{d-3}{2a}\right) n =
-2 \pi\, i\,T_H\,n,
\end{equation}
so that the d-dimensional Hawking temperature increases linearly
with the number of extra dimensions: $T_H=(d-3)/(4\pi a)$.

\section{Final Comments}

In this paper we have used the Fock-Ivanenko formalism for the
Dirac equation in curved spacetime to write the Dirac equation for
a massless $1/2$-spin-fermion around a Schwarzschild BH. We have
calculated approximated analytical solutions for the spinors. The
radial eigenfunctions of the spinors
$\phi^{(\uparrow\downarrow)}_l$ are described by Schr\"odinger
-like equations for the radial eigenfunctions of the spinors,
which can be rewritten in terms of the redefined tortoise
coordinate $u(r)$: $v=-(u-a)/a$. We have proven the SUSY character
of the potentials $V^{(\uparrow\downarrow)}_1$ and
$V^{(\uparrow\downarrow)}_2$, corresponding to the Hamiltonians
${\rm H}^{(\uparrow\downarrow)}_1$ and ${\rm
H}^{(\uparrow\downarrow)}_2$, which are not hermitian. We obtained
approximated analytical solutions for eigenvalues of energy (or
quasi-normal frequencies). The interesting result here obtained is
that all the asymptotic states (and their SUSY partners) have
purely imaginary frequencies as eigenvalues of energy, which can
be expressed in terms of the Hawking temperature $T_H$:
$E^{(\uparrow\downarrow)}_n=-2\pi\,i\,n T_H$. Furthermore, the
isolated bottom state has a real null eigenvalue of energy, in
agreement with what one expects for a SUSY
Hamiltonian\cite{nogami}. Therefore, all the asymptotic modes
decay, except the whole with zero energy. The treatment with
$d>4$- compact extra dimensions deserves a more detailed study and
goes beyond the scope of this work. However, some comments were
included in Sect. VII.

\acknowledgments{ The authors acknowledge CONICET and UNMdP
(Argentina) for financial support.}

\begin{appendix}
\section{Quantization of quasi-normal frequencies: states}\label{ap1}

We consider the equation (\ref{sch}) for the massless Dirac modes
$\Psi^{(\uparrow\downarrow)}_{1,2}$. The case with
$\kappa_{\uparrow}=(\ell+1)$ accounts for the
$\Psi^{(\uparrow)}_{1,2}$, (with $\ell=0, 1, 2, ...$), while
$\kappa_{\downarrow}=-\ell$ corresponds to the
$\Psi^{(\downarrow)}_{1,2}$ ones (of course with $\ell=1, 2,
...$).

Although two elections are possible for the potentials (the second
exponential term in the potential can take signs $+$ or $-$), the
Schr\"{o}dinger equation itself seems to come from a Morse-like
potential problem. In symbols:
\begin{eqnarray}
  -\frac{d^{2}\Psi_{1}^{(\uparrow\downarrow)}}{d\upsilon^{2}}+\{4(\kappa_{\uparrow\downarrow})^{2}e^{-2\upsilon}
  +2(\kappa_{\uparrow\downarrow}) e^{-\upsilon}-4a^{2}[E_{n}^{(1,\uparrow\downarrow)}]^{2}\}\Psi_{1}^{(\uparrow\downarrow)} &=& 0, \\
  -\frac{d^{2}\Psi_{2}^{(\uparrow\downarrow)}}{d\upsilon^{2}}+\{4(\kappa_{\uparrow\downarrow})^{2}e^{-2\upsilon}
  -2(\kappa_{\uparrow\downarrow}) e^{-\upsilon}-4a^{2}[E_{n}^{(2,\uparrow\downarrow)}]^{2}\}\Psi_{2}^{(\uparrow\downarrow)} &=& 0.
\end{eqnarray}

Then, we take the way followed by Morse to solve our
problem\cite{M}. First, we make a change of variables:
\begin{equation}
  y = e^{-\upsilon},   \quad \frac{d}{d\upsilon} = -y\frac{d}{dy},
  \quad
  \frac{d^{2}}{d\upsilon^{2}} =
  y\frac{d}{dy}+y^{2}\frac{d^{2}}{dy^{2}}.
\end{equation}
The resultant equations are
\begin{eqnarray}
  \frac{d^{2}\Psi_{1}^{(\uparrow\downarrow)}}{dy^{2}}+\frac{1}{y}\frac{d\Psi_{1}^{(\uparrow\downarrow)}}{dy}
  +\{\frac{[\xi_{n}^{(1,\uparrow\downarrow)}]^{2}}{y^{2}}-\frac{2(\kappa_{\uparrow\downarrow})}{y}
  -4(\kappa_{\uparrow\downarrow})^{2}\}\Psi_{1}^{(\uparrow\downarrow)} &=& 0, \\
  \frac{d^{2}\Psi_{2}^{(\uparrow\downarrow)}}{dy^{2}}+\frac{1}{y}\frac{d\Psi_{2}^{(\uparrow\downarrow)}}{dy}
  +\{\frac{[\xi_{n}^{(2,\uparrow\downarrow)}]^{2}}{y^{2}}+\frac{2(\kappa_{\uparrow\downarrow})}{y}
  -4(\kappa_{\uparrow\downarrow})^{2}\}\Psi_{2}^{(\uparrow\downarrow)} &=& 0,
\end{eqnarray}
where the parameters $4a^{2}[E_{n}^{(i,\uparrow\downarrow)}]^{2}$
have been replaced by $[\xi_{n}^{(i,\uparrow\downarrow)}]^{2}$.
Finally, a clever transformation of functions bring us closer to a
problem with hypergeometric differential equations. Let the set of
functions be $F^{(i,\uparrow\downarrow)}(y)$ such that
\begin{equation}\label{fun1}
    \Psi_{1}^{(\uparrow\downarrow)}(y) = e^{-\alpha y}\,(2\alpha
    y)^{\frac{\beta}{2}}\,F_{1}^{(\uparrow\downarrow)}(y),
\end{equation}
\begin{equation}\label{fun2}
    \Psi_{2}^{(\uparrow\downarrow)}(y) = e^{-\alpha y}\,(2\alpha y)^{\frac{\beta}{2}}\,F_{2}^{(\uparrow\downarrow)}(y).
\end{equation}
At the same time we define the variable $z$ as
\begin{equation}
  z = 2\alpha y, \quad  \frac{d}{dy} = 2\alpha \frac{d}{dz}, \quad  \frac{d^{2}}{dy^{2}} = 4\alpha^{2}\frac{d^{2}}{dz^{2}}.
\end{equation}
Putting it all together, we arrive at the final expression, which
entails particular cases of the general confluent hypergeometric
differential equation\cite{Abramowitz}
\begin{eqnarray*}
   z \frac{d^{2}F_{1}^{(\uparrow\downarrow)}}{dz^{2}}&+&(1+\beta-z)\frac{dF_{1}^{(\uparrow\downarrow)}}{dz}
  \nonumber \\
  & + &
  \left\{\frac{4[\xi_{1}^{(\uparrow\downarrow)}]^{2}+\beta^{2}}{4z}-\frac{2(\kappa_{\uparrow\downarrow})
  + \alpha(\beta+1)}{2\alpha}+\frac{\alpha^{2}-4(\kappa_{\uparrow\downarrow})^{2}}{4\alpha^{2}}z\right\} F_{1}^{(\uparrow\downarrow)} = 0, \\
  z
  \frac{d^{2}F_{2}^{(\uparrow\downarrow)}}{dz^{2}}&+&(1+\beta-z)\frac{dF_{2}^{(\uparrow\downarrow)}}{dz}\nonumber
  \\
   &+&  \left\{ \frac{4[\xi_{2}^{(\uparrow\downarrow)}]^{2}+\beta^{2}}{4z}+\frac{2(\kappa_{\uparrow\downarrow})-\alpha(\beta+1)}{2\alpha}
  + \frac{\alpha^{2}-4(\kappa_{\uparrow\downarrow})^{2}}{4\alpha^{2}}z \right\} F_{2}^{(\uparrow\downarrow)} = 0.
\end{eqnarray*}
Under certain conditions (which we shall study later), these
equations have the form of a confluent hypergeometric differential
equation
\begin{equation}
    z\frac{d^{2}F_{n\beta}(z)}{dz^{2}}+(1+\beta-z)\frac{dF_{n\beta}(z)}{dz}+n\,F_{n\beta}(z)=0.
\end{equation}
The general solution can be expressed in terms of the Laguerre
polynomials ${\rm L}(n,\beta,z)$
\begin{equation}
    F_{n\beta}(z)=A_{n\beta}\,{\rm L}(n,\beta,z) +B_{n\beta} \,\left[\frac{\Gamma(\beta)}{\Gamma(-n)}\frac{{\rm L}(n+\beta,-\beta,z)}{{\rm bin}(n,\beta) \, z^{\beta}}
    +\frac{\Gamma(-\beta)}{\Gamma(-n-\beta)}\frac{{\rm L}(n,\beta,z)}{{\rm bin}(n+\beta,n)} \right],
\end{equation}
where ${\rm bin}(n, \beta) = n!/\left[\beta!(n-\beta)!\right]$.
Whatever the rest of the arguments, if $n$ becomes positive
integer, the solution always will be able to be written as a
polynomial form (finite number of terms). This treatment of the
solutions is particularly useful for the states; and we have the
ground states $(n=0)$ using the fact that $F_{0\beta}=const.$ in
the attractive potentials.

There are four significant cases of interest, which we shall study
separately.

\subsubsection{$\kappa_{\uparrow}=\ell+1$ and
$F_{1}^{(\uparrow)}$}\label{a1}

The first case is
\begin{eqnarray}
  4[\xi_{n}^{(1,\uparrow)}]^{2}+\beta^{2} &=& 0, \\
  -\frac{2(\kappa_{\uparrow})+\alpha(\beta+1)}{2\alpha} &=& n, \\
  \alpha^{2}-4(\kappa_{\uparrow})^{2} &=& 0.
\end{eqnarray}
A decaying exponential factor in (\ref{fun1}) forces us to set
$\alpha=2(\ell+1)$; then we obtain $\beta=-2(n+1)$. Finally
$\xi_{n}^{(1,\uparrow)}=i\frac{\beta}{2}=-i(n+1)$. The minus
election in the sign of $\xi$ is because of the energy
\begin{equation}
    E_{n}^{(1,\uparrow)}=\frac{\xi_{n}^{(1,\uparrow)}}{2a}=-i\frac{(n+1)}{2a}
\end{equation}
which must have a negative imaginary part to recover the outgoing
quasi-normal modes. Notice that $n$ must be a positive integer,
and due to the fact $\beta=-2(n+1)<0$, we obtain $n=0,1,2,3,...$

\subsubsection{$\kappa_{\uparrow}=\ell+1$ and
$F_{2}^{(\uparrow)}$}\label{a2}

In this case
\begin{eqnarray}
  4\left[\xi_{n}^{(2,\uparrow)}\right]^{2}+\beta^{2} &=& 0, \\
  \frac{2(\kappa_{\uparrow})-\alpha(\beta+1)}{2\alpha} &=& n', \\
  \alpha^{2}-4(\kappa_{\uparrow})^{2} &=& 0.
\end{eqnarray}
Using similar arguments, but with (\ref{fun2}), we obtain
$\alpha=2(\ell+1)$, $\beta=-2n'$ and
$\xi_{n}^{(2,\uparrow)}=i\frac{\beta}{2}=-in'$. Again $n'$ must be
a positive integer.
\begin{equation}\label{energ}
    E_{n'}^{(2,\uparrow)}=\frac{\xi_{n'}^{(2,\uparrow)}}{2a}=-i\frac{n'}{2a}.
\end{equation}
Using the fact that $\beta=-2n'<0$, we obtain $n'=1,2,3,...$, so
that $n=n'-1$, and the energies (\ref{energ}) can be written as
\begin{equation}
E_{n}^{(2,\uparrow)}= E_{n}^{(1,\uparrow)}=-i\frac{(n+1)}{2a},
\end{equation}
where $n=0,1, ...$

\subsubsection{$\kappa_{\downarrow}=-\ell$ and
$F_{1}^{(\downarrow)}$}

Only the main results are of interest for us;
      $\alpha=2\ell$, $\beta=-2m$ and
      $\xi_{m}^{(1,\downarrow)}=i\frac{\beta}{2}=-im$, with $m$ a
      positive integer. The energy in this case is
\begin{equation}
    E_{m}^{(1,\downarrow)}=\frac{\xi_{m}^{(1,\downarrow)}}{2a}=-i\frac{m}{2a},
\end{equation}
where $m=1,2,...$

\subsubsection{$\kappa_{\downarrow}=-\ell$ and $F_{2}^{(\downarrow)}$}

In this case $\alpha=2\ell$, $\beta=-2(m'+1)$ and
$\xi_{m}^{(2,\downarrow)}=i\frac{\beta}{2}=-i(m'+1)$, with $m'$ a
positive integer. The energy in this case is
\begin{equation}
    E_{m'}^{(2,\downarrow)}=\frac{\xi_{m'}^{(2,\downarrow)}}{2a}=-i\frac{(m'+1)}{2a}.
\end{equation}
A similar analysis to those made in (\ref{a1}) and (\ref{a2}),
gives us $m'=m-1$ and finally we find that the energies (or
quasi-normal frequencies) have a unique result (we relabel $m
\rightarrow n$)
\begin{equation}
E_{n}^{(1,\downarrow)}=E_{n}^{(2,\downarrow)}=-i\frac{n}{2a},
\end{equation}
where $n=1,2,...$.
\end{appendix}

\end{document}